\documentclass[twocolumn,prd,nofootinbib,longbibliography,eqsecnum]{revtex4-2}
\pdfoutput=1
\usepackage{amsmath}
\usepackage{booktabs}
\usepackage{amssymb}
\usepackage{mathrsfs}
\usepackage{bm}
\usepackage{graphicx}
\usepackage{color}
\newcommand{\beq}{\begin{equation}}
\newcommand{\eeq}{\end{equation}}

\def\apjs{Astrophys.\ J.\ Supp.}
\def\apjl{Astrophys.\ J.\ Lett.}

\def\jcap{J.\ Cosmol.\ Astropart.\ Phys}

\begin{document}

\title{Simulating Partial Sky Cosmic Microwave Background Maps with 3D Fast Fourier Transforms}
\author{Mariona Giner Mascarell and Emory F. Bunn}
\affiliation{Physics Department, University of Richmond, Richmond, VA  23173, USA}

\begin{abstract}
Simulated maps of the microwave background (CMB) radiation are generally created using one of two methods: all-sky simulations use the spherical harmonic transform, while maps covering small areas approximate the sky as flat, allowing the use of fast Fourier transforms (FFTs). Current and near-future experiments, particularly ones like CMB S4, will cover areas too large for the flat-sky approximation but significantly less than the full sky. In this regime, it can be more efficient to simulate maps in a 3-D box using FFTs, and then sample onto the observed part of the celestial sphere. We present a method for performing such simulations and show that it can be more efficient than full-sky simulations. We develop the method for scalar maps, but we expect it to be applicable to higher-spin (e.g., polarization) simulations as well. 
\end{abstract}

\maketitle

\section{Introduction} \label{Introduction}

Maps of the comic microwave background (CMB) radiation are among the most important data sets in cosmology. Beginning with the COBE DMR data \cite{cobe}, temperature anisotropy maps covering several orders of magnitude in angular scale have provided powerful cosmological constraints \cite{challinor}, and more recently CMB polarization maps have begun to do the same.

In order to make full use of these maps, it is often necessary to compare observations with simulations. We typically model the CMB as a realization of a homogeneous and isotropic Gaussian random process on the sphere. Efficient tools for simulating such processes are therefore essential. 

Two approaches are typically taken for such simulations. The precisely correct method is to simulate the entire sky using spherical harmonic transforms, as implemented most famously in the HEALPix software \cite{healpix}. The time required for such simulations is $O(N_{\mathrm{pix}}^{3/2})$ where $N_{\mathrm{pix}}$ is the number of pixels on the entire sky. 
On the other hand, when comparing with observations that cover only a small fraction of the sky, one instead models the observed patch as flat, allowing for simulation via fast Fourier transforms (FFTs) (e.g., \cite{numericalrecipes}) rather than spherical harmonic transformations.

Historically, most experiments have fit neatly into one category or another: satellites \cite{cobe,wmap,planck} require full-sky maps, while suborbital experiments (e.g., \cite{spt3g,sptpol,actpol,bicep}) cover areas for which the flat-sky approximation is appropriate. But as the sky coverage of the latter category of experiment grows, a middle ground, more precise than the flat sky but more efficient than the spherical harmonic transform,  becomes more desirable. This need may be expected to become more acute with the development of instruments similar to CMB S4 \cite{cmbs4}.

The CMB can be regarded as a specific realization of a random Gaussian field characterized by its mean and covariance properties, which encapsulate the statistical homogeneity and isotropy of the primordial perturbations from which it originated (e.g., \cite{dodelson}). In particular, because of the small amplitude of the density perturbations over the relevant part of the Universe's history, the observed temperature fluctuations are linearly related to the perturbations in the density, gravitational potential, etc., all of which may be modeled as homogeneous and isotropic Gaussian random processes. 
%For this reason, generating a CMB anisotropy map is in principle a straightforward task.  The observed temperature fluctuation in any direction is derived from 

%derived from can be described as a function of position whose Fourier transform is a set of complex numbers. This means that we can simulate a sky map with the right statistical properties by choosing Gaussian random numbers for each entry in the Fourier transform. In this case, the standard deviations of each distribution are given by the variance of the entries in the Fourier transform, also known as power spectrum $P(k)$, which is a function of the wavenumber $k$. 

%This proves to be highly effective for simulating a map of the microwave background on a square patch of the sky. However, there is a limitation on the size of the sky patch that can be accurately represented, as Fast Fourier Transforms (FFTs) are defined in Euclidean space. When the patch of sky becomes too large to utilize FFTs effectively, spherical harmonics can be employed as an alternative to handle the curvature of the celestial sphere. Mathematically, these harmonics are expressed as 

The observed CMB temperature fluctuation may be expanded in spherical harmonics,
\begin{equation}
    T(\theta, \phi) = \sum_{l=0}^{\infty} \sum_{m=-l}^{l} \Tilde{T}_{lm} Y_{lm} (\theta, \phi),
\label{eq:temperature}
\end{equation}
where $Y_{lm} (\theta, \phi)$ are the spherical harmonics and $\Tilde{T}_{lm}$ are their coefficients. Under the above assumptions, the coefficients $\tilde T_{lm}$ are independently  drawn from complex normal distributions with mean zero and variances $\langle |\tilde T_{lm}|^2\rangle\equiv C_l$, where the angular power spectrum $C_l$ may be derived from the assumptions of the theory under consideration (e.g., the power spectrum of the primordial gravitational potential, the matter content, etc.).

If we are interested in a sufficiently small patch of the sky, we can instead model the temperature anisotropy with a 2-D Fourier transform:
\begin{equation}
    T(\vec r) = \sum_{\vec k}\tilde T_{\vec k}e^{i\vec k\cdot\vec r},
\end{equation}
in which the vectors $\vec r, \vec k$ are two-dimensional and the coefficients $\tilde T_{\vec k}$ are drawn from complex normal distributions with mean zero and variances $P(k)$ that are related in a straightforward way to the angular power spectrum $C_l$. 

Because the exact spherical harmonic computation is relatively computationally expensive, it is natural to seek methods for simulating the CMB sky that have the advantages of FFTs but can be applied over regions too large for the flat-sky approximation. In this paper, we examine an approach to this problem based on simulating the function $T$ in a three-dimensional box that contains the observed portion of the celestial sphere. To be specific, for any given angular power spectrum $C_l$, we seek a 3-D power spectrum that produces the same two-point correlation function over a large enough range of separations. We can then use a 3-D FFT to simulate in that box and sample the resulting map onto the sphere. Even if the observed region is too large for the flat-sky approximation to be acceptable, the box may be quite small in one of the three dimensions, as we will see.

Note that the 3-D power spectrum need not be derived from a physical model; rather, it simply must match the correlations produced by the true angular power spectrum over the relevant range of separations. Because a Gaussian random process is completely determined by its one- and two-point functions, maps produced by this method will have the same statistical properties as maps produced via a spherical harmonic transform.

Whether this method is more efficient than the spherical harmonic transformation depends on the fraction of sky covered and the required angular resolution. As we will show, for an interesting range of these parameters it is possible to get improvements in efficiency.

An alternative to this approach is to seek ways to make the spherical harmonic transform more efficient \cite{tiansoft,tianapjs,reinecke,reinecke2,suda,tygert1,tygert2,slevinsky,driscoll,mohlenkamp,drake}. These include approaches that have better asymptotic scaling than the $O(N^{3/2})\log N$ scaling of the HEALPix algorithm but have numerical stability issues or prove in practice not to be more efficient for realistic cases. Others (e.g., \cite{tiansoft,tianapjs,reinecke,reinecke2} have similar scaling but achieve efficiencies through greater parallelization, use of GPUs, etc.

The remainder of this paper is structured as follows. Section \ref{Formalism} lays out the mathematical formalism behind the problem. Section \ref{testing_the_box} presents the results of tests of whether the maps are indeed statistically equivalent to the spherical-harmonic approach. Section \ref{results} presents results of our tests, and Section \ref{sec:discussion} provides a brief discussion. We explored one variant of the method that proved not to be of great use as currently implemented but may become so with further refinement; we describe this method in an Appendix.

\section{Formalism} \label{Formalism}

Our method uses FFTs to simulate in a three-dimensional box that contains the points of the sky that we want to measure. In order to provide these simulations with the right statistical properties, we must find a three-dimensional power spectrum for the box. Then, we can use it to simulate maps in the box and later sample these onto the sphere. Now, we delve into the specifics of each step, outlined in their respective subsections.

%Each subsection provides detailed instructions on how to perform these individual steps.

\subsection{The Power Spectrum In the Box}

Our starting point is to find a power spectrum for the box that has the same two-point correlation function as that corresponding to the angular power spectrum $C_l$.  The relationship between the angular power spectrum and the two-point correlation function in the sphere $\xi_{\textrm{sph}}$ is

\begin{equation}
    \xi_{\mathrm{sph}}(\theta) = \sum_l^\infty \frac{(2l+1) C_l P_l (\cos\theta)}{4 \pi},
\label{eq:harmonics}
\end{equation}
where $P_l (\cos\theta)$ are the  Legendre polynomials. If we wish to make maps that have been smoothed, we use the appropriately smoothed angular power spectrum. For instance, for smoothing with a Gaussian beam of width $b$, we use
\begin{equation}
    C_{l}^{\mathrm{smooth}} = C_l \cdot e^{-b^2 l^2}.
\label{eq:cl}
\end{equation}

Taking the celestial sphere to have radius 1, any two points on the sphere separated by an angle $\theta$ are separated by a distance \begin{equation}
    r=2\sin(\theta/2)
    \label{eq:rtheta}
\end{equation}
in three-dimensional space. We want to find a power spectrum $P(k)$ for a Gaussian random process in three-dimensional space whose correlation function $\xi(r)$ corresponds to $\xi_{\mathrm{sph}}(\theta)$.
 
 Although we will eventually simulate in a finite-sized box, we begin by finding an appropriate power spectrum for infinite flat space. We will then then restrict it to the dimensions of the box. 
 
 We recall that the correlation function $\xi_\infty$ in flat space is the Fourier transform of the power spectrum $P_\infty$. Writing the Fourier integral in spherical coordinates and integrating over the angles leads to the standard result
 \begin{equation}
    \xi_{\infty}(r) = 4\pi\int_0^{\infty} k \, P_{\infty}(k) \, \frac{\sin(k \, r)}{r} \, dk.
    \label{eq:xiinfpinf}
\end{equation}

\begin{figure*} 
    \centerline{
    \includegraphics[width=7.6in]{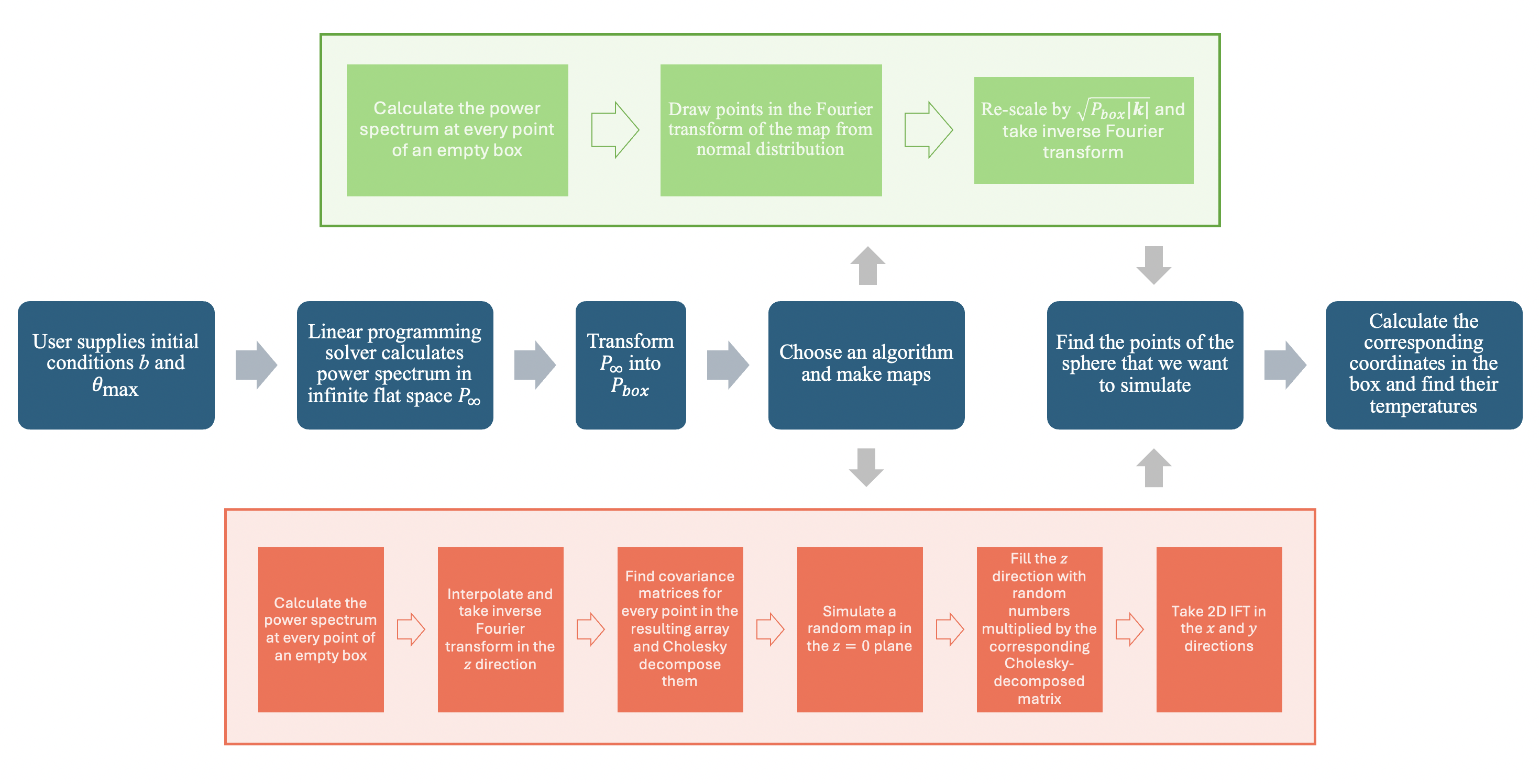}} 
    \caption{Flowchart for the code illustrating the chronology of the computational process. The green boxes represent the 3-D Fourier method and the orange boxes represent the 2-D Fourier  method.}
    \label{flow}
\end{figure*}

\subsection{Linear Programming}
\label{sec:linprog}
We wish to find a power spectrum $P_\infty$ whose correlation function (\ref{eq:xiinfpinf}) 
satisfies $\xi_\infty(r) = \xi_{\mathrm{sph}}(\theta)$, where $r$ and $\theta$ are related by equation (\ref{eq:rtheta}), out to a given maximum separation $\theta_{\mathrm{max}}$.
This is a set of linear constraints on the values of $P(k)$.  
In addition, we must insist that the power spectrum be nonnegative for all $k$.

After discretizing in both $k$ and $r$ to make the number of constraints and unknowns finite, this is a
standard linear programming problem. To be specific, we phrase the above task as a problem that involves finding a vector $\bm{x}$ that minimizes a linear objective function $\bm{c}^T\bm{x}$, subject to the constraints $\bm{A} \bm{x} \geq \bm{b}$ and $\bm{x} \ge 0$. In this case, the vector that we need to find is a discrete version of $P_{\infty}$, the power spectrum in infinite flat space. On the other hand, $\bm{b}$ is the vector containing the correlation function at a discretized set of values of $r$ up to some maximum $r_{\mathrm{max}}$.
The matrix $\mathbf{A}$ is then performs a discretized version of the integral in equation (\ref{eq:xiinfpinf}).

We in fact wish to impose equality constraints $\bm{Ax}=\bm{b}$, but we find it convenient to impose inequality constraints instead, and then choose the objective function such that it is minimized when all the constraints are equalities (assuming such a solution to exist). To be specific, this corresponds to choosing the elements of the vector $\bm{c}$ so that $\bm{c}^T\bm{x}$ is the sum of the elements of $\mathbf{Ax}$, by choosing
\begin{equation}
    c_j = \sum_iA_{ij}.
\label{eq:sumconstraint}
\end{equation}
The objective function is $\bm{c}^T\bm{x} = \sum_j(\bm{Ax})_j$.
The inequality constraints require that this quantity is always greater than or equal to $\sum_jb_j$. A solution in which this inequality is an equality will be one in which $\bm{Ax}=\bm{b}$ as desired, and if such a solution exists, it will be the minimum found by solving the linear programming problem.

%The matrix $A$ will be the discrete matrix form representation of Equation (2.3). In other words, $A$ is an operator that transforms a given power spectrum into the corresponding correlation function. Furthermore, since we are imposing an inequality constraint rather than an equality constraint, our linear objective function to be minimized is $\sum_k (A\cdot \bm{x})_k$. 

Because we are discretizing both the power spectrum and the correlation function, we need to specify some parameters. First, we define $\theta_{\textrm{max}}$ as the maximum polar angle of the region of the sphere that we are interested in simulating. The maximum separation between two points on the spherical cap being simulated is then  $r_{\textrm{max}} = 2 \cdot \sin(\theta_{\textrm{max}})$.  We impose constraints on the correlation function at a set of points between 0 and $r_{\mathrm{max}}$ separated by an amount $\Delta r$. so the number of constraints (i.e., the dimension of $\mathbf{b}$) is ${r_{\textrm{max}}}/{\Delta r}$. Similarly, we define $k_{\textrm{max}}$ as the value after which we assume that the power spectrum is zero, and $\Delta k$ as the separation between $k$ values.  Thus, the power spectrum will be described by a vector $\mathbf{x}$ with $k_{\textrm{max}}/{\Delta k}$ entries. 

When using this algorithm, the user supplies the angular power spectrum $C_l$, along with values for the smoothing parameter $b$ and the desired maximum polar angle $\theta_{\textrm{max}}$. The rest of the parameters are specified in terms of these. In particular, $\Delta r$ needs to be smaller than the smoothing parameter to capture small features, so we define it as $\Delta r ={b}/{5}$.  Likewise, $\Delta k$ needs to be smaller than the minimum $k$ value that fits in the box. So we define it as $\Delta k = {2 \pi}/({3 r_{\textrm{max}}})$.

\subsection{Making the Box}

Once the power spectrum in infinite flat space is obtained by solving the linear programming problem, we can start constructing the box. The first natural thing to do is deciding how big we want the box to be. We choose the number of points that we want in every direction, as well as the separation between them. The box will have an equal width and length, each with $N$ points, and a smaller height with $M$ points. The distance between points $\Delta p$ is constant in every direction, and we define it as $0.85 \Delta r$ to make sure we able to capture fine details.

Of course, the specific values of $N$ and $M$ will depend on the size of the part of the sphere that we want to simulate. In other words, they will depend on $\theta_{\textrm{max}}$. 
%Moreover, we need to make sure that they are big enough to accommodate all the necessary waves. 
For $N$, we began by doubling the length of the part of the sky that we want to simulate, and then rounding it up to the next power of two, since that  makes FFTs work better. For $M$, we doubled the height of the part of the sky that we want to simulate, and again, rounded it up to the next power of two. We later added a variable in front, called the size factor, to conveniently test different sizes of the box. 

Once we have decided on a size for the box, we can finally calculate its power spectrum by narrowing it down to the sizes of our box. This is straightforward, again, if we remember that the correlation function and the power spectrum are Fourier transforms of each other in flat space. Approximating this relationship as a sum, we obtain 
\begin{equation}
    \xi (\bm{r}) = \sum_k P_{\infty}({k}) e^{-i \bm{k} \cdot \bm{r}}  \, \delta k_x \,  \delta k_y \, \delta k_z .
\label{eq:discxidels}
\end{equation}
Note that we can express the power spectrum in the box as a Fourier series, 
\begin{equation}
    \xi (\bm{r}) = \sum_k P_{\textrm{box}}(k ) e^{-i \bm{k} \cdot \bm{r}}.
\label{eq:pboxxi}
\end{equation}
Therefore, 
\begin{equation}
    P_{\textrm{box}} = P_{\infty} \, \delta k_x \,  \delta k_y \, \delta k_z,
\label{eq:pboxandinf}
\end{equation}
 where
 %$\delta k_x$, $\delta k_y$, and $\delta k_z$ are the separations in each direction between points sampled in the discrete form of the Fourier transforms. To decide on some values for these, we must remember that the waves of the Fourier transform that fit into the box will have wavenumber $k = \frac{2 \pi}{\lambda}$, where $\lambda$ is the wavelength. In this case, the wavelength is the size of the box, which is $N \Delta p$ for the width and length and $M \Delta p$ for the height. Therefore, 
\begin{equation}
    \delta k_x = \delta k_y = \frac{2 \pi}{N \Delta p}
\label{eq:smalldeltakxy}
\end{equation}
and
\begin{equation}
    \delta k_z = \frac{2 \pi}{M \Delta p}.
\label{eq:smalldeltakz}
\end{equation}
Thus, the relationship between the power spectrum in infinite flat space and the power spectrum in the box is 
\begin{equation}
    P_{\textrm{box}} = P_{\infty} \left( \frac{2 \pi}{\Delta p} \right)^3 \frac{1}{M N^2} .
\label{eq:finalpbox}
\end{equation}

\begin{figure*} 
    \centerline{
    \includegraphics[scale=0.35]{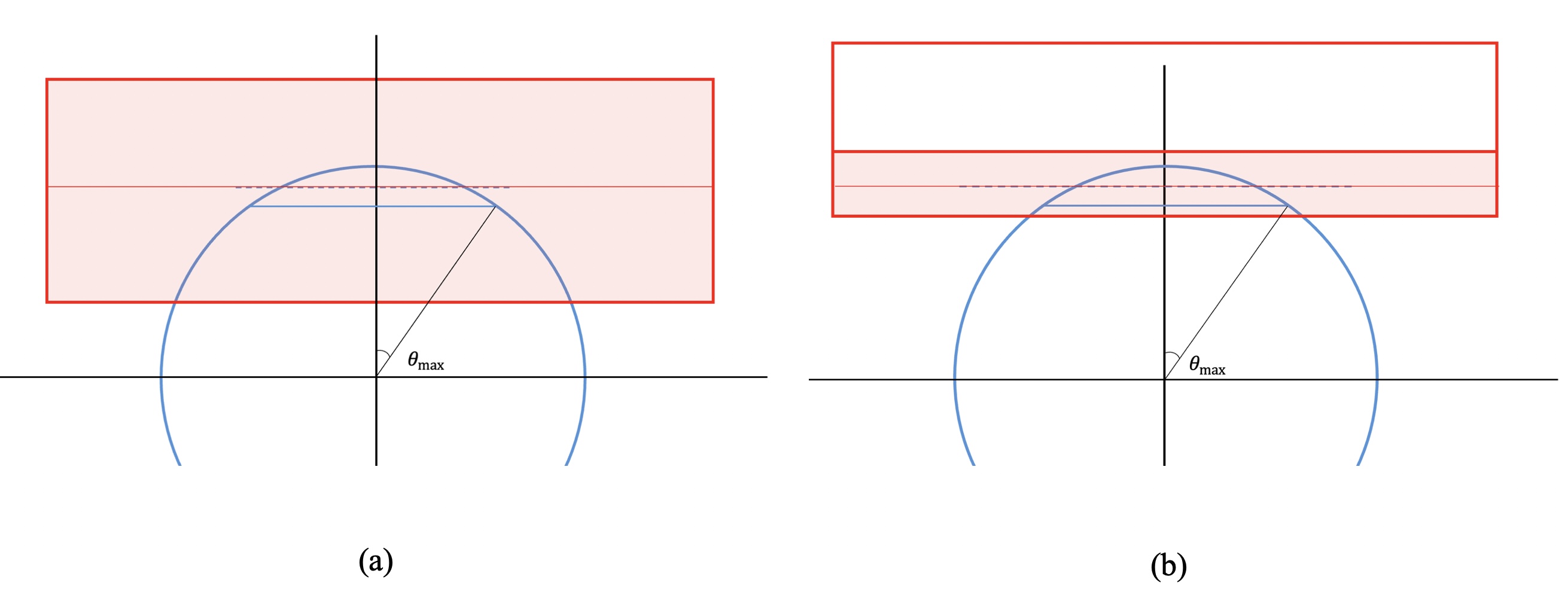}} 
    \caption{Two-dimensional sketch of the relative positions of the box and sphere for the 3-D Fourier method (a) and the 2-D Fourier method (b). }
    \label{boxes}
    \label{fig:boxsphere}
\end{figure*}

\subsection{Simulating in the Box}

Once we have obtained the power spectrum in the box, we can begin simulating maps. First, we will simulate in the entire box and then we will sample that onto the sphere. Specifically, we developed two different algorithms to simulate the maps, each aimed for different sizes of the box in order to minimize running time. The steps of the two methods are shown in Fig. \ref{flow}.

\subsubsection{The 3-D Fourier Method}

The first method is the more straightforward algorithm. In this approach, we create a box of dimensions $N \times N \times M$, populate it with simulated values of the 3-D Fourier transform $\tilde T_{\vec{k}}$, and apply an inverse FFT to simulate a map. This case is especially fast when the smoothing parameter is big, leading to small box sizes.

To be specific, we determine $k=\sqrt{k_x^2+k_y^2+k_z^2}$ for each point in the $N\times N\times M$ box, and populate each entry in the box with a normally-distributed complex random number with mean zero and variance $P(k)$. We then take the inverse Fourier transform of this array, giving a complex function $T(x,y,z)$ with the desired correlations. The real and imaginary parts of this function are then two independent realizations of the appropriate 3-D random process in the box.

%In order to start simulating in this box, we consider the Fourier transform of a CMB map. Even though every point in the Fourier transform is drawn from a different normal distribution, for coding purposes we first create an array full of numbers that come from identical distributions. It is important to remember that the power spectrum gives the variance of the map, that is, the square of the standard deviation of the distribution every point was drawn from. Therefore, we can re-scale the numbers with the square root of the power spectrum for every value of $\bm{k}$. Because we are working in a three-dimensional space, $\bm{k}$ is a vector, but we can calculate the power spectrum of its magnitude. That is, the re-scaling of the map will be done using the factor
%\begin{equation}
%    \sqrt{P_{\textrm{box}}(|\bm{k}|)}
%\label{eq:sqrtpbox}
%\end{equation}
%The only thing left to do at this point is taking the inverse Fourier transform, thus obtaining a CMB map that fills the entire box.

Once we have simulated in the whole box, it is time to sample the function at the points corresponding to pixels on the sphere. To do so, we first must interpolate the box map. Then, for any given $\theta_{\textrm{max}}$, we can find the locations of all of the pixels of interest (which, for a spherical cap centered on the pole, will be the first pixels in a HEALPix map stored in RING mode). 

 %One thing to take into account is that the box and the sphere have different coordinate systems. The sphere where we want to simulate is a unit sphere, but the box is $N \Delta p$ long and wide, and $M \Delta p$ tall, which means that their scales are different. Because we need to find the points in the box that correspond to points in the sphere, we need to find some coordinate conversion rules. 

The coordinates of the HEALPix pixels are most naturally represented as cartesian coordinates $(x,y,z)$ for points on the unit sphere, with $(0,0,1)$ being the pole. We take the $z$ axis of our box to be the same as that of the sphere, and center the box at coordinate $(1+z_{\mathrm{min}})/2$, where $z_{\mathrm{min}}=\cos^{-1}(\theta_{\mathrm{max}})$ is the $z$ coordinate of the edge of the spherical cap, as shown in Fig. \ref{fig:boxsphere} (a).

%Since the box and the sphere are centered, their $z$-axes overlap. Now, the correct placement of the box becomes very important. For the \change{old method}, the box is sliced in half in every direction by its axes, so that the width and length go between $-\frac{N}{2}$ and $\frac{N}{2}$ and the height goes from $-\frac{M}{2}$ to $\frac{M}{2}$.  The key is to place the box so that its $x$-axis overlaps with the horizontal line cutting the region of interest of the sphere by half, as depicted in Figure 2(a). Thus, the $x$ and $y$ coordinates of the box are the same as those from the sphere, only that they need to be re-scaled by $\Delta p$. Now, the box and the sphere have their $z$-axes aligned but they are shifted. With basic trigonometry, we find that 
%\begin{equation}
%    B_z = \frac{S_z}{\Delta p} - \left( \frac{\cos \theta}{\Delta p} + \left( \frac{1 - \cos\theta}{2 \Delta p}\right) \right)
%\label{eq:coordsystem}
%\end{equation}
%where $B_z$ is the $z$-coordinate in the box and $S_z$ is the $z$-coordinate in the sphere.
%Once we have located the desired points in the box, we can calculate the temperature at every found coordinate. 

\subsubsection{The 2-D Fourier Method}

For some choices of parameters, we may be able to achieve greater efficiency by performing simulations in Fourier space in the $x$ and $y$ directions but in real space in the $z$ direction. The reason for this is that the number $h$ of layers of the box needed for the simulation in the $z$ direction is often quite small ($h\ll M$). Simulating in real space in the $z$ direction means that we no longer have the $M\log M$ scaling of the FFT, but the smaller value of $h$ can compensate for the poorer asymptotic scaling properties of the method.

To be specific, we simulate our box maps in a hybrid space: Fourier space in the $x$ and $y$ directions and real space in the $z$ direction.
Let us call these quantities $\hat T(k_x,k_y,z)$. We need to simulate these numbers in a box of size $N\times N\times h$ (as opposed to   $N\times N\times M$ in the 3-D Fourier method). 
Once we have generated these quantities, we do an inverse FFT in the $x$ and $y$ directions to get the real-space box simulation. 

In this approach, the random numbers we must generate are independent for pairs of points with different values of  $k_x$ and $k_y$ dimensions, but for any given $(k_x,k_y)$, there are correlations between points at different heights $z$. To find these correlations, we populate an $N\times N\times M$ box with the values of the power spectrum $P_{\mathrm{box}}(k)$, and perform an FFT in the $z$ direction. Because of the well-known fact that the correlation function and the power spectrum are a Fourier transform pair, the resulting values give the two-point correlation function 
\begin{equation}
    \xi_{k_xk_y}(z_1-z_2)\equiv \langle\hat T(k_x,k_y,z_1)\hat T(k_x,k_y,z_2)\rangle.
\end{equation}

The specific procedure to simulate these correlated random numbers is as follows. For each $(k_x,k_y)$, we generate the $h\times h$ covariance matrix $\bm\Gamma$ of the values of $\hat T$: 
\begin{equation}
\Gamma_{jk} = \langle \hat T(k_x,k_y,z_i)\hat T(k_x,k_y,z_j)\rangle=\xi_{k_xk_y}(z_i-z_j). 
\end{equation}
%The values of this matrix are sampled directly from the Fourier-transformed power spectrum described above. 
We can now simulate an $h$-dimensional random vector by the following standard procedure: we Cholesky decompose $\bm\Gamma$,
\begin{equation}
    \bm{\Gamma} = \mathbf{L} \cdot \mathbf{L}^\top
\label{eq:gammacholesky}
\end{equation}
and apply $\mathbf{L}$ to an $h$-dimensional vector of independent standard Gaussian random numbers.

Note that the matrices $L$ for all $k_x,k_y$ may be precomputed and used to generate many simulations.

Having generated these $h$-dimensional random vectors for all $k_x,k_y$, all that remains is to perform the 2-D FFT.

For this approach, it is conceptually nicer to think of the coordinate systems for the sphere and box as defined by the right panel of Fig. \ref{fig:boxsphere}. That is, we place the center of the simulated region in the middle of the region of height $h$, rather than in the middle of the larger $M$-dimensional box (which is never actually generated). 

%With the Cholesky decomposition, we can transform a list of independent (uncorrelated) random numbers $\bm{x}$ into a list of correlated numbers $\bm{m_z}$ that adhere to a specified covariance matrix. With the calculation of these matrices concludes the first part of the algorithm, which only needs to be computed once for every batch of maps.

%Next, for every point in this two-dimensional space, we fill the $z$ direction by generating independent random numbers drawn from normal distributions and multiplying them by their corresponding Cholesky-decomposed covariance matrices, which were calculated previously. Lastly, we take the 2D inverse Fourier transform in the $x$ and $y$ directions, thus ending up with a CMB map that is equivalent to one created with the first method, but with less layers in the $z$ direction. The last thing to do is to sample these points onto the sphere. The procedure is the same as the one for the old method, the only difference being the placement of the box. In this case, the part of the sphere that we want to simulate is centered with the part of the box that contains the simulations, as seen in Figure 2(b).

It is worth noting that a Cholesky decomposition can be viewed as a matrix analog to taking the square root. That is why any alternative matrix factorization that is also analogous to a square root should in principle result in equivalent results. Since any covariance matrix is symmetric by definition, a reasonable alternative to Cholesky is an eigendecomposition. This method factorizes $\bm\Gamma$ as
\begin{equation}
    \bm{\Gamma} = \mathbf{P D P}^{-1}
\label{eq:gammasvd}
\end{equation}
where $\mathbf{P}$ is an orthogonal matrix whose columns are the eigenvectors of $\bm{\Gamma}$ and $\mathbf{D}$ is a diagonal matrix whose entries are the corresponding eigenvalues. Thus, $\mathbf{PD}^{1/2}$ would be analogous to the Cholesky decomposition. In fact, this factorization proved to be more numerically stable, so it is the one that we adopted to test the new method.

\section{Tests} \label{testing_the_box}

Having explained the details of the two algorithms, we now turn to the methods used for testing their accuracy and reliability. Both algorithms were evaluated using the same tests.
We begin by describing some simple initial tests and then proceed to the most thorough test.
\begin{figure*} 
    \centerline{
    \includegraphics[scale=0.4]{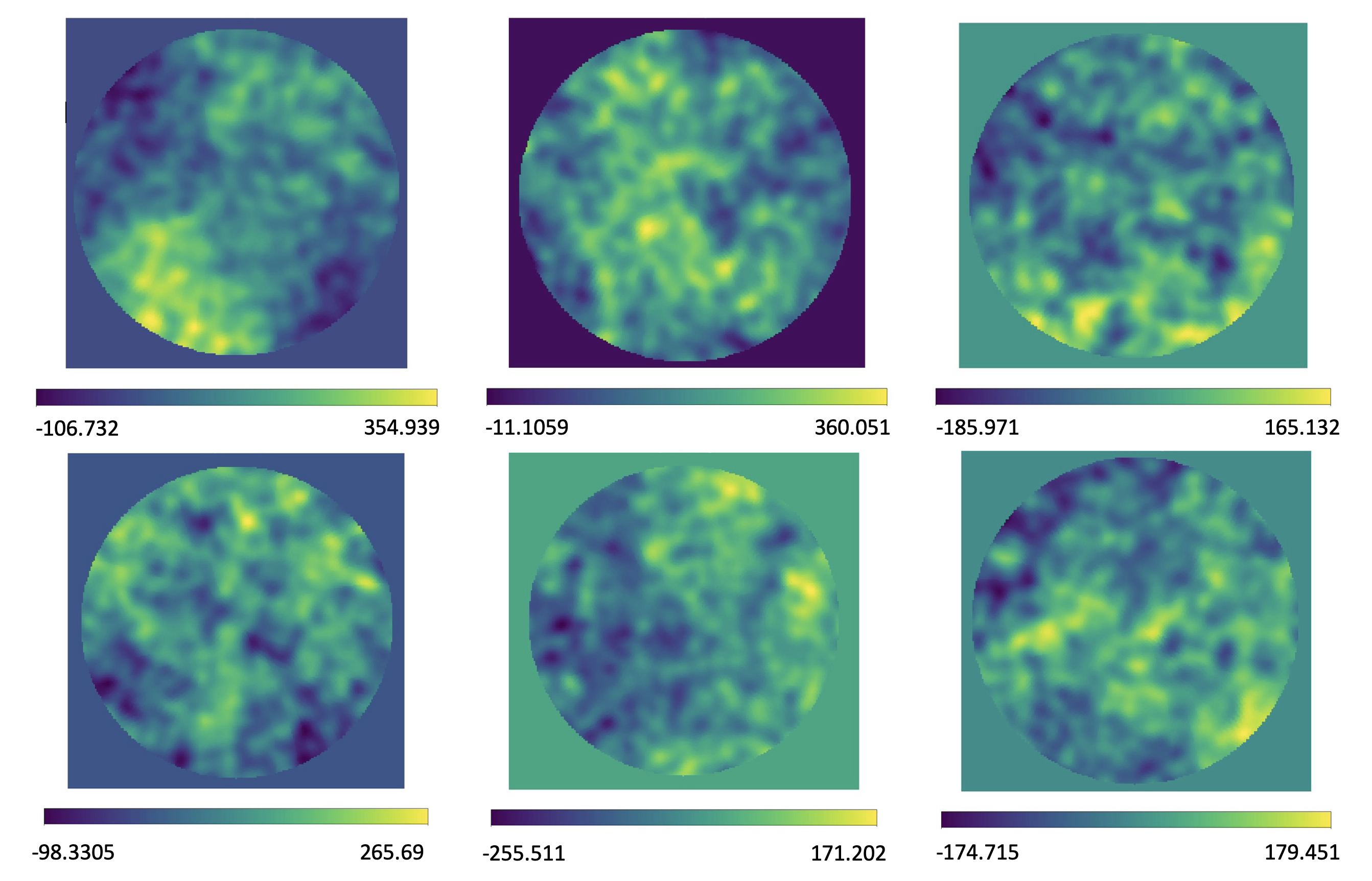}} 
    \caption{The top row shows maps made with the Box Method and the bottom row shows maps made with Healpy. For both, the smoothing parameter is $b=0.1$ rad ($0.57^\circ$), the polar angle is $\theta_{\mathrm{max}}=15^\circ$, and the units are $\mu K$.}
    \label{fig:samplemaps}
\end{figure*}

\subsection{Initial tests}
\label{initialtests}
\paragraph{The Correlation Function At Zero Separation}
As mentioned before, our goal is to produce maps with the correct two-point correlation function. Because all random numbers we generate are Gaussian with zero mean and all transformations are linear, our resulting maps will be samples from a Gaussian random process with zero mean. Such a process is completely described by its two-point function, so as long as our method produces correct two-point correlations, it is correct. A simple but important first test is to keep track of the two-point correlations at zero separation, $\xi(0)$, to make sure it remains the same through the various stages of the computation.

The ``true'' value of $\xi(0)$ is found by evaluating equation (\ref{eq:harmonics}) at $\theta=0$. 
We can compare with the results of  equations (\ref{eq:xiinfpinf}) and (\ref{eq:discxidels}) at $r=0$ to check the solution $P_\infty$ of the linear programming problem its discretization $P_{\mathrm{box}}$.
Moreover, because $\xi(0)=\langle T^2\rangle$, we can compare these results with the mean-square values of our simulations in the box and on the sphere.

%the power spectrum in flat space calculated with the linear programming is supposed to have the same correlation function as the angular power spectrum. This is how we make sure that the box-produced maps have the right statistical properties. Specifically, it is crucial that the value of the correlation function at zero separation, $\xi (0)$, is maintained throughout the steps in the algorithm. Thus, by plugging $r=0$ into equations 2.1, 2.3, and 2.5, we were able to check along the way whether this value was conserved for the sphere, infinite flat space, and the box, respectively. An important property of the correlation function at zero separation is that it represents the variance of the map. In other words, $\xi (0)$ for a given map is the mean of the squares of each temperature point. This allowed us, once we had our final box-produced map, to check whether this value was still conserved in our final step.

\paragraph{The pseudo-$C_l$ Test}
We wish to compare our maps with those generated by HEALPix. One natural arena in which to do this is to compute the angular power spectra of maps generated by HEALPix with those from our method, to see if the estimated $C_l$ coefficients have the same statistical properties.

Because our maps cover only part of the sky, we compare them with HEALPix-generated maps with the same sky coverage, as well as the same input angular power spectrum. We use HEALPix's {\tt anafast} routine to estimate the power spectra for both sets of maps and compare the averages. Because we use maps with partial sky coverage, the output of {\tt anafast} consists of ``pseudo-$C_l$s'' rather than the true power spectrum, but since our goal is to compare the two methods with each other, this does not present a problem.

Because the statistical properties of the pseudo-$C_l$s are not simple, it is not obvious how to make this a precise statistical test, but it is powerful nonetheless: errors in the algorithm are likely to lead to quite noticeable visual discrepancies.

%During the development of the Old and New algorithms, HEALPix (Hierarchical Equal Area isoLatitude Pixelization) played a crucial role by enabling us to make meaningful comparisons of maps made with Healpy (the corresponding Python package) with those produced with the box. Beyond that, Healpy has a function called "anafast" that calculates the angular power spectrum given a map. To do so, they employ spherical harmonics, so this function is technically only applicable to full-sky maps. However, we used it for partial-sky maps made both with Healpy and the box intending to potentially detect some discrepancies between the two. We acknowledge that this isn't valid proof that the box-produced and healpy-produced maps are statistically equivalent and that the results are difficult to interpret, but it did help us draw conclusions about the range of applicability of both the Old and New method.
\begin{table*}[t]
\centering
\scriptsize
\renewcommand{\arraystretch}{1.3}
\setlength{\tabcolsep}{6pt}
\resizebox{\textwidth}{!}{%
\begin{tabular}{cccccccccc}
\toprule
\textbf{$b$} (deg) & \textbf{$\theta_{\max}$} (deg) & \textbf{$N$} & \textbf{$M$} & \textbf{$h$} & \textbf{$N_{\text{side}}$} & \textbf{Noise \%} & \textbf{Time Factor} & \multicolumn{2}{c}{\textbf{Healpix Time Factor}} \\
\midrule
0.573  & 15   & 256   & 16   & 4  & 64  & 0.5  & $\mathbf{4.2\times10^{6}}$   & \multicolumn{2}{c}{$1.1\times10^{7}$} \\
0.573  & 20   & 256   & 32   & 5  & 64  & 0.5  & \textbf{$\mathbf{5.2\times10^{6}}$}   & \multicolumn{2}{c}{$1.1\times10^{7}$} \\
0.573  & 25   & 256   & 32   & 6  & 64  & 0.5  & \textbf{$\mathbf{6.3\times10^{6}}$}   & \multicolumn{2}{c}{$1.1\times10^{7}$} \\
0.573  & 30   & 512   & 64   & 7  & 64  & 0.5  & $3.3\times10^{7}$            & \multicolumn{2}{c}{$1.1\times10^{7}$} \\
0.286 & 15   & 512   & 32   & 5  & 128 & 0.36 & \textbf{$\mathbf{2.4\times10^{7}}$}   & \multicolumn{2}{c}{$8.7\times10^{7}$} \\
0.286 & 20   & 512   & 64   & 7  & 128 & 0.47 & \textbf{$\mathbf{3.3\times10^{7}}$}   & \multicolumn{2}{c}{$8.7\times10^{7}$} \\
0.286 & 25   & 512   & 64   & 9  & 128 & 0.6  & \textbf{$\mathbf{4.2\times10^{7}}$}   & \multicolumn{2}{c}{$8.7\times10^{7}$} \\
0.286 & 30   & 1024  & 128  & 12 & 128 & 0.7  & $3.4\times10^{8}$            & \multicolumn{2}{c}{$8.7\times10^{7}$} \\
\bottomrule
\end{tabular}
} % end resizebox
\caption{Comparison of the 2-D Fourier method with HEALPy. We mark in boldface those cases where the time factors indicate that the box method is faster than HEALPy.}
\label{table:mainresults}
\end{table*}

\subsection{Cholesky Decomposition and Kolmogorov - Smirnov test}

Our method purports to produce vectors that are drawn from a multivariate normal distribution with mean zero and known covariance matrix $\Gamma$. To be specific, the covariance between any two pixels in our maps is given by equation (\ref{eq:harmonics}), where $\theta$ is the angular separation between the two pixels.

We adopt the following procedure  to test whether our maps are consistent with this hypothesis. We first find the Cholesky decomposition $\mathbf{L}$ of $\bm{\Gamma}$. Then, for each simulated map $\bm{m}$, we compute $\bm{x}=\bm{L}^{-1}\bm{m}$. If our maps have the desired covariances, then $\bm{x}$ should be a vector of independent standard Gaussian random variables. We then apply a Kolmogorov-Smirnov test to see if these values are consistent with this model.

%Another very important property that characterizes the CMB is that any map should be a list of correlated random numbers following a covariance matrix $\Gamma$. Each entry in this covariance matrix indicates the level of correlation between the two pixels corresponding to the indexes of that entry. Since this correlation can be easily calculated using Equation 2.1, it's relatively straightforward to calculate the covariance matrix using the smoothed $C_l$. To check whether the box-produced maps have the desired statistical properties, we use a Cholesky decomposition. In this case, the vector $\bm{x} = L^{-1} \bm{m}$, where $\bm{m}$ is the box-produced map, should be a list of uncorrelated random numbers with a mean of 0 and a standard deviation of 1.

%To verify whether this is true, we applied the Kolmogorov-Smirnov (KS) test to determine if $\bm{x}$ was actually composed of uncorrelated random numbers. However, 
Unfortunately, the covariance matrix $\Gamma$ is not well conditioned, having many eigenvalues near zero. As a result, numerical errors render
the covariance matrix non-positive definite, preventing the Cholesky decomposition from being performed. To address this, varying amounts of noise were added to the covariance matrix to ensure it became positive definite, and a corresponding amount of Gaussian white noise was added to the maps $\bm{m}$, before performing the KS test. 

This test may be considered decisive. If the box-produced maps pass it with sufficiently small noise levels, we conclude that the covariance properties of the simulated maps are correct. As noted above, since the maps are made from linear processes applied  are mathematically guaranteed to be Gaussian with zero mean. As a result, the covariances are the only thing that must be tested.

\section{Results} \label{results}

In this section, we present the results of testing the 2-D Fourier Method using a series of standardized test cases. The tests were conducted on batches of 1000 maps that ran on the University of Richmond's Spydur computing cluster. Specifically, we tested different values of the smoothing parameter, maximum polar angles, and sizes of the box.

While it is important to run quantitative tests on the box-produced maps, Fig. \ref{fig:samplemaps} provides visual evidence that the 3-D Fourier method produces CMB maps that are visually indistinguishable from those made with HEALPy. This was in fact true for all the combinations of parameters that we tested.

As discussed in Section \ref{testing_the_box}, the Kolmogorov-Smirnov (KS) test is the strongest test of consistency between the maps made with the box method and those made with HEALPy. After visual inspection and using the initial tests in Section \ref{initialtests}, we use the KS test as the final arbiter of success of the method.

With no noise added to the maps, the box method consistently fails the KS test, but adding small amounts of noise typically remedies the problem. This occurs because the two methods differ in the typical amplitudes of certain modes with very low signal levels. Adding low levels of white noise to both simulations masks this difference, making the methods statistically indistinguishable. 

When simulating experimental data, the need to add small amounts of noise does not present a problem, as such simulations will have larger amounts of noise added anyway. In applications for which one does not anticipate adding noise, further tests may be required to see if the need to add noise affects the results in important ways.

Table \ref{table:mainresults} summarizes the results of simulations performed for various smoothing parameters $b$ and map sizes $\theta_{\mathrm{max}}$. In each case, the dimensions of the box ($N,M,h$) are shown, along with the HEALPix size $N_{\mathrm{side}}$. In each case, we show the noise level required to pass the KS test. 

We also show a comparison of two ``time factors'' to indicate how the two methods may be expected to compare in speed. For the box method, the slowest step is the set of $h$ two-dimensional Fourier transforms, each of dimension $N\times N$. We therefore define the time factor for the box method to be
\begin{equation}
    \mbox{Box Time Factor}  = 
    N^2\log_2(N^2)h
\end{equation}
based on the standard scaling of the FFT.\footnote{Populating the $N\times N\times h$ box takes time $N^2h^2/2$. For the cases we consider, this term is always smaller.}

Simulation of a HEALPix map, on the other hand, scales as the number of pixels to the 3/2 power:
\begin{equation}
    \mbox{HEALPix Time Factor} = N_{\mathrm{pix}}^{3/2}=(12N_{\mathrm{side}}^2)^{3/2}.
\end{equation}

Table \ref{table:mainresults} shows both of these factors. Naturally, in both cases, the actual scaling times have prefactors of order 1. As a result, these values should be compared in order of magnitude rather than precisely. Nonetheless, we highlight in bold the cases where the box time factor is smaller, suggesting that the box method may be faster.

We do not present results of a direct timing comparison between the two methods because we have not attempted at this stage to optimize the box code, which is written entirely in Python. We leave to future work the task of creating optimized code in a compiled language for a fair end-to-end comparison of the methods. All we claim at the moment is that our results show that for a significant range of parameters, the box method may prove faster.

\section{Discussion}
\label{sec:discussion}

We have examined the possibility of simulating the CMB sky using 3D FFTs in a box rather than spherical harmonic transforms. We have demonstrated that the method can work in principle. To be specific, it is possible to find a 3D power spectrum $P$ that reproduces the same correlations as the angular power spectrum $C_l$ for realistic models. Moreover, we have shown that the method can produce maps that are nearly indistinguishable from maps produced via the spherical harmonic transform, with only small amounts of noise required to make them completely indistinguishable. 

For some choices of sky coverage and angular resolution, the method may prove to be faster than the spherical harmonic transformation.

A natural next step in this work is to code the box method more efficiently, in a compiled language, so that we can replace the scaling calculations in Table \ref{table:mainresults} with a fully realistic comparison. In addition, there are multiple avenues to try to improve the method, including further attempts to get the ``shrinking'' method described in the Appendix to work.

We have focused in this work on scalar (temperature) CMB maps, but there is no reason the method cannot be generalized to higher-spin maps. As the interest in CMB studies shifts further toward polarization maps (e.g., \cite{cmbpol-review}), a spin-2 generalization of this method would be of particular interest.

%Originally, we made simulations only with a size factor of $1$ (see section II.C). However, preliminary results from the pseudo-$C_l$ test suggested that those sizes for the box were not enough to accommodate all necessary waves to provide the maps with fine structure. Thus, we repeated all the simulations with new size factors of $8$ and $16$, whose pseudo-$C_l$ tests led us to believe were big enough. As shown in Table I, using these bigger size factors also increases the accuracy of the values of the correlation function at zero separation. Based on these findings, we chose 16 as our standard size factor. 

%We perfomed the Kolmogorov-Smirnov test on the different algorithms, for different values of the smoothing parameter, polar angle, and sizes of the box. To present the results, we used the 2-D Fourier method. See Table I.

%\section{Conclusions} %\label{conclusions}

\section*{Acknowledgements}
We gratefully acknowledge the support of the University of Richmond School of Arts \& Sciences, which provided summer research fellowships to both authors.

\bibliography{Bibliography}
\begin{table*}[t]
\centering
\scriptsize
\renewcommand{\arraystretch}{1.3}
\setlength{\tabcolsep}{6pt}
\resizebox{\textwidth}{!}{%
\begin{tabular}{cccccccccc}
\toprule
\textbf{$b$}(deg) & \textbf{$\theta_{\max}$}(deg) & \textbf{$N'$} & \textbf{$M$} & \textbf{$h$} & \textbf{$N_{side}$} & \textbf{$s$} & \textbf{Noise \%} & \textbf{Time Factor} & \textbf{Healpix Time Factor} \\
\midrule
0.573 & 15  & 256  & 128  & 5  &  64  & 0.71 & 1.4\% & $\mathbf{5.2\times10^{6}}$ & $1.1\times10^{7}$ \\
0.573 & 20  & 256  & 256  & 5  &  64  & 5    & 1.8\% & $\mathbf{5.2\times10^{6}}$ & $1.1\times10^{7}$ \\
0.573 & 25  & 256  & 256  & 6  & 64  & 10   & 2.4\% & $\mathbf{6.3\times10^{6}}$ & $1.1\times10^{7}$ \\
0.573 & 30  & 512  & 256  & 7  & 64  & 5    & 3.0\% & $\mathbf{3.3\times10^{7}}$ & $1.1\times10^{7}$ \\
0.286 & 15  & 512  & 256  & 5  &  128  & 0.4  & 1.3\% & $\mathbf{2.4\times10^{7}}$ & $8.7\times10^{7}$ \\
0.286 & 20  & 512  & 512  & 7  & 128  & 1.5  & 1.5\% & $\mathbf{3.3\times10^{7}}$ & $8.7\times10^{7}$ \\
0.286 & 25  & 512  & 512  & 9  & 128  & 6    & 5.2\% & $\mathbf{4.2\times10^{7}}$ & $8.7\times10^{7}$ \\
0.286 & 30  & 1024 & 1024 & 12 & 128  & 15   & 4.0\% & $2.5\times10^{8}$ & $8.7\times10^{7}$ \\
\bottomrule
\end{tabular}%
}
\caption{Simulation parameters and corresponding time scales for the shrinking method.}
\label{table:shrinking}
\end{table*}

\section{Appendix} \label{Appendix}

An alternative method was developed in order to improve time efficiency. This is not presented in the results since, despite proving to be faster than HEALPy for a wider range of angles, it proved to be less accurate, requiring significantly more noise to pass the KS test.

We call this method the Shrinking Method, since it consists of shrinking the box in order to reduce the time for the 3D FFT. We will now sketch the idea behind this method.

We wish to find a power spectrum $P$ that reproduces the correlation function $\xi$ up to some maximum value $r_{\mathrm{max}}$. In order to find a valid solution to the linear programming problem, we typically have to choose a box whose size is much larger than $r_{\mathrm{max}}$.

We will describe the concept in one dimension since its generalization to higher dimensions is straightforward. 

Suppose that we have found such a solution using discrete Fourier transforms (DFT) of size $N$. That is, we have found a discretized power spectrum $P(k)$ at $N$ points separated by an interval $\Delta k$ whose discrete Fourier transform matches the desired correlation function for values of $r<r_{\mathrm{max}}$. We take the DFT of the power spectrum to get the correlation function. We then truncate it to some size $N'<N$, which is still large enough to contain the entire constrained region, and Fourier transform back to get a new power spectrum of dimension $N'$. By construction, this power spectrum will produce the correct correlation function over  the constrained region. We can then use this power spectrum in a smaller box (of size $N'$ rather than $N$). 

In order for this to work, the power spectrum after shrinking must be positive everywhere. This will not generally be true, but if it is, we have found a smaller box, and hence improved the efficiency of the algorithm.\footnote{One might think that, if such a positive power spectrum of smaller dimension exists, we could have found it by simply solving the linear programming problem with a smaller box to begin with. This would be true in one dimension but is not necessarily true in more than one. The reason is that we solve the linear programming problem for the power spectrum $P_\infty$ assuming that it is isotropic -- that is, that it is a function of only the magnitude of $\vec k$. The shrinking happens after we use $P_\infty$ to populate the values of $P_{\mathrm{box}}$. The shrinking process may  produce a box power spectrum that cannot be derived from an isotropic $P_\infty$ and hence could not have been found by the linear programming algorithm. }

%Given that the unconstrained parts are not of primary interest, we can truncate them. Thus, by restricting the correlation function to only smaller values of $r$, we obtain a correlation function for a smaller box that still matches in the constrained region. Then, we can Fourier transform back and obtain the power spectrum in the smaller box. The size of the new box is $N' \times N' \times M$ where $N'$ is calculated the same way as $N$ but with a smaller size factor.

In order for this method to produce a positive power spectrum, we generally need to introduce one more step. Before giving the correlation function to the linear programming solver, we divide it by a Gaussian function of $r$. After solving the linear programming problem, we  convolve the  power spectrum with the Fourier transform of this Gaussian, so that the final power spectrum will produce a correlation function that matches the correct constraints.

The reason this works is the power spectrum found by solving the linear programming problem is typically very spiky. The truncation step is equivalent to convolving the power spectrum with an oscillating function, so the narrow spikes in the untruncated power spectrum have negative values. The convolution step smooths out the sharp peaks in the power spectrum, reducing this problem.

%This is because, otherwise, the power spectrum that the linear programming returns is very spiky. Then, when the shrinking step is applied, this causes the existence of significant negative values in the smaller box power spectrum. In order to prevent this, we divide our correlation function by a Gaussian $f$ and through the linear programming solver, we obtain a ``fake" power spectrum. Then, we can convolve this fake power spectrum with the Fourier transform of the Gaussian function $\hat{f}$ to obtain a smooth true power spectrum in the box.

To be specific, we divide the original correlation function constraints by
%Specifically, the Gaussian function used to divide the correlation function was 
\begin{equation}
f(r) = \left( \sqrt{\frac{2 \pi}{s}}\right)^3 \cdot e^{-{r^2}/({2 s^2})}
\end{equation}
for some constant $s$, which  must be chosen to be small enough that a solution to the linear programming problem exists.
We then solve the linear programming problem with these constraints, finding a solution $P_{\mathrm{fake}}$ Then, the true power spectrum is
\begin{align}
    P_{\textrm{true}} &= \int P_{\mathrm{fake}}(\bm{k'}) \hat{f}(\bm{k} - \bm{k'}) \, d^3 k' \nonumber \\
    &= \int_0^\infty dk' \, P_{\mathrm{fake}}(k') \, k'^2 \int d\Omega' \, \hat{f}(|\bm{k} - \bm{k'}|) \nonumber \\
    &= \int_0^\infty dk' \, P_{\mathrm{fake}}(k') \, k'^2 \cdot W(\bm{k}, \bm{k'}),
\end{align}
where 
\begin{equation}
    W(\bm{k}, \bm{k'}) = \frac{2 \pi \bm{k'}}{s^2 \bm{k}} ( e^{\frac{-1}{2} s^2 (\bm{k}-\bm{k'})^2} -  e^{\frac{-1}{2} s^2 (\bm{k}+\bm{k'})^2}).
\end{equation}

%While this convolution process significantly reduces negative values, additional measures are required to completely eliminate them. For instance, this is also the reason why we added the supplementary constraints detailed in Section \ref{sec:linprog}. After applying these measures, the negative values become sufficiently small, allowing us to effectively set them to zero. 

%For the larger angles and finer resolution parameters, these negative values in the power spectrum become significant, leading to large values of noise to pass the KS test, and hence making this method, as of right now, not competitive. 
Unfortunately, although we were able to reduce the negative values in the shrunken power spectrum, the method did not eliminate them entirely. As a result, larger amounts of noise were required to produce simulations that passed the KS test.

Table \ref{table:shrinking} shows the results of applying this method. Although the shrinking results in substantial improvements in speed, the required noise levels are high enough to make the method less useful in many applications. It may be possible to find variations on this method that reduce the noise levels.

\end{document}